\documentclass[10pt, a4paper]{article}

\usepackage{mathtext}        
\usepackage[T2A]{fontenc}
\usepackage[cp1251]{inputenc}
\usepackage[english,russian]{babel}
\usepackage{courier}
\usepackage{indentfirst}     
\usepackage[dvips]{graphicx}
\usepackage{amsmath}
\usepackage{amssymb}
\usepackage{wrapfig}
\usepackage{textcomp}         

\usepackage{listings}

\newcounter{myfig}
\newcommand{\ash}{\mathop{\mathrm{ash}}\nolimits}

\makeindex
\begin{document}

\begin{center}
{\Huge\bf Являются ли жесткие неинерциальные системы отсчёта жесткими?}

\vskip 10mm

{\huge С.\,С.\,Степанов\footnote{e-mail: phys@synset.com}} \\
~\\

\vskip 5mm

\parbox{14cm}{
\large
В работе анализируется понятие жесткости системы отсчёта в рамках специальной теории относительности.
Сформулированы три определения жесткости.
На примерах различных неинерциальных систем продемонстрирована их  неэквивалентность.
Показано, что  так называемые  жесткие неинерциальные системы отсчёта Мёллера  обладают локальной
жесткостью,  но не являются жесткими в глобальном смысле.
Обсуждается физическая причина этого явления и её связь
со смыслом неевклидовости геометрии пространства в неинерциальной системе отсчёта.
}

\end{center}

\Large

\section{Введение}

В этой работе мы сформулируем три,  на первый взгляд эквивалентных,  определения  жесткости.
На примерах различных систем отсчёта  будет показано, что на самом деле они не совпадают.
В частности,  класс произвольно ускоренных вдоль прямой систем отсчёта Мёллера традиционно называют жесткими.
Тем не менее,  будет продемонстрировано, что они не являются жесткими для любых
двух удалённых наблюдателей этой системы.

Статья организована следующим образом.
Сначала мы введём  некоторые базовые понятия, которые потребуются в дальнейшем.
Затем сформулируем определения жёсткости системы отсчёта.
Чтобы продемонстрировать их неэквивалентность,  мы рассмотрим четыре
неинерциальные системы.
В последней из них, двигающейся прямолинейно с произвольной переменной скоростью,
выяснится, что локальная жёсткость системы
отсчёта  не влечет за собой жесткости глобальной.
Мы обсудим физическую причину этого явления и её связь со смыслом неевклидовости геометрии
3-пространства в неинерциальных системах отсчёта.
Будет доказано, что единственной системой,  в которой эти типы жесткости совпадают
в случае поступательного движения, является жесткая равноускоренная система отсчёта Борна-Мёллера.
В приложении в ковариантном виде рассмотрены исходные рассуждения Борна, при помощи
которых он в 1909 г. сформулировал свой критерий жёсткости.

\newpage

\section{Неинерциальные системы отсчёта}

Метрика пространства Минковского в лабораторной инерциальной системе отсчета $S_0:\,\{T,\,X,\,Y,\,Z\}$ имеет вид:
\begin{equation}\label{ds_mikowskij}
ds^2 =dT^2 - dX^2-dY^2-dZ^2,
\end{equation}
где $T$ -- физическое время, измеряемое синхронизированными часами, $X^i=\{X,\,Y,\,Z\}$ -- декартовы координаты события.
Мы используем систему единиц,   в которой скорость света равна единице.
Греческие индексы изменяются от 0 до 3, а латинские от 1 до 3.

Произвольную систему отсчёта $S:\,\{t,\,x,\,y,\,z\}$ удобно определять, задавая законы движения каждой её точки
относительно лабораторной системы  $S_0$.
Пусть $x^i=\{x,\,y,\,z\}$ -- координаты, однозначно фиксирующие данную точку системы $S$. Эта точка движется
относительно лабораторной системы $S_0$ по траектории:
\begin{equation}\label{traject_NIRF}
         X^i(T)=F^i(T,\,x,\,y,\,z).
\end{equation}
Время $t$ неинерциальной системы можно определить любым удобным способом при помощи  произвольной функции $T = T(t,\,x,\,y,\,z)$.
Предполагается только, что более ранние события в $S$ соответствуют меньшим значениям $t$,
чем более поздние. Такое время является координатным и,  вообще говоря,  не совпадает с физическим временем
часов,  связанных с точкой $x^i$.
Заменяя в траектории (\ref{traject_NIRF}) время $T$ на  $T(t,\,x,\,y,\,z)$, мы получаем преобразования
от системы $S$ к лабораторной системе $S_0$:
\begin{equation}\label{transf_gen}
T = T(t,\,x,\,y,\,z),~~~~~~~~X^i=X^i(t,\,x,\,y,\,z).
\end{equation}
Подстановка этих преобразований в (\ref{ds_mikowskij}) даёт
интервал между событиями в неинерциальной системе отсчёта:
\begin{equation}\label{ds_mikowskij_nonin}
ds^2 =g_{\alpha\beta}\,  dx^\alpha dx^\beta,
\end{equation}
где метрика $g_{\alpha\beta}$ автоматически обладает нулевой кривизной
(в специальной теории относительности 4-пространство  псевдоевклидово).

В рамках данной системы отсчёта всегда можно перейти к другому способу нумерации событий:
\begin{equation}\label{transf_in_this_system}
         t' = t'(t,\,x,\,y,\,z),~~~~~~~~x'^i=x'^i(x,\,y,\,z).
\end{equation}
Первое преобразование определяет новое координатное время, а оставшиеся -- другой способ нумерации
пространственных точек системы. Важно, что последние не зависят от времени $t$, т.к., в противном случае
мы бы получили другую систему отсчёта.

\newpage

$\bullet$ Чтобы при выбранном способе нумерации событий $\{t,\,x,\,y,\,z\}$ делать некоторые физические заключения,
необходимо определить как эти координаты связаны с физическими временем $\delta \tau$ и длиной $\delta l$.
Распишем интервал (\ref{ds_mikowskij_nonin}), отделив нулевой индекс ($x^0=t$):
\begin{equation}
ds^2 =g_{00}\,dt^2+2g_{0i}\,dtdx^i + g_{ij}\,dx^i dx^j
\end{equation}
и выделим полный квадрат по координатному времени $dt$:
\begin{equation}
ds^2 =\left(\sqrt{g_{00}} \,dt + \frac{g_{0i}}{\sqrt{g_{00}}}\, dx^i\right)^2 - \left(-g_{ij}+\frac{g_{0i}g_{0j}}{g_{00}}\right)\, dx^idx^j.
\end{equation}
Это выражение имеет псевдоевклидовый вид
$
ds^2 = \delta \tau^2-\delta l^2,
$
где
\begin{equation}\label{nonint_physT}
\delta \tau =  \sqrt{g_{00}} \,dt + \frac{g_{0i}}{\sqrt{g_{00}}}\, dx^i
\end{equation}
называется {\it физическим временем}.
Если $dx^i=0$, то  физическое время совпадает с собственным временем часов $\delta \tau_0 =  \sqrt{g_{00}} \,dt$, находящихся в точке $x^i$.
В общем случае $\delta \tau$ равно времени прохождения светового сигнала в одну сторону между двумя соседними точками,
часы в которых локально синхронизированы \cite{Stepanov2013}.
Если $\delta \tau$ оказывается полным дифференциалом, то в такой системе отсчета можно ввести единое для всех
точек (наблюдателей) синхронизированное время $\tau=\tau(t,\,x,\,y,\,z)$.

Элемент бесконечно малой {\it физической длины}:
\begin{equation}\label{nonint_physL}
\delta l^2 = \gamma_{ij}\, dx^idx^j,~~~~~~~~~~~~\gamma_{ij} =-g_{ij}+\frac{g_{0i}g_{0j}}{g_{00}}
\end{equation}
имеет смысл {\it радиолокационного расстояния},  которое получает наблюдатель,
расположенный в точке пространства с координатами $x^i$ до бесконечно близкой к нему точки $x^i+dx^i$ \cite{Landau2_2003}.
Для его измерения он посылает сигнал со скоростью света,  который,  отражаясь от точки $x^i+dx^i$, возвращается обратно.
Время движения такого сигнала в обе стороны равно удвоенному расстоянию $\delta l$ между точками.
Выражение (\ref{nonint_physL}) можно также получить \cite{Myelller1987}, рассматривая сопутствующую
инерциальную систему отсчёта,  в которой точка $x^i$ в данный момент времени имеет нулевую скорость.
Эталоны длины и времени наблюдателя такой  системы,
совпадают с аналогичными эталонами неподвижного относительно него неинерциального наблюдателя.

Величины  $\delta \tau$ и $\delta l$ инвариантны относительно преобразований (\ref{transf_in_this_system}),
а,  следовательно,  не зависят от способа нумерации событий  в данной системе отсчёта.
Распространение света соответствует нулевому интервалу $ds=0$ и его скорость всегда равна единице:  $\delta l/\delta \tau=1$,
а координатная скорость света $dx^i/dt$ может быть и больше единицы.

\newpage

\section{Определения жесткости системы отсчёта}

Жесткость (как бы мы её не определяли) -- это понятие относительное.
``Парадокс Белла'' является хорошей иллюстрацией к этому утверждению.
Пусть  две точки движутся так, что расстояние между ними остаётся неизменным относительно лабораторной системы отсчёта.
Относительность одновременности приводит к тому,
что для наблюдателей, связанных с каждой точкой, это же расстояние  зависит от времени.
Аналогично, если расстояние между наблюдателями
в неинерциальной системе неизменно, то оно может оказаться зависящем от времени с точки зрения лабораторной
системы отсчёта.

Поэтому, обсуждая далее жесткость системы отсчёта,  мы будем подразумевать,  что она выполняется для
наблюдателей, связанных с точками этой системы.
Аналогично собственному времени, будем называть её {\it собственной жесткостью} системы отсчёта.

Возможны по крайней мере три определения собственной жесткости:

\vskip 3mm

\begin{quote}
I. {\it Сопутствующая жесткость}: все точки системы отсчёта имеют нулевую скорость в сопутствующей к ней инерциальной системе отсчёта.
\end{quote}

\vskip 3mm

\begin{quote}
II. {\it Локальная жесткость}: тензор $\gamma_{ij}=-g_{ij}+g_{0i}g_{0j}/g_{00}$, определяющий элемент бесконечно малой
физической длины,  не зависит от времени.
\end{quote}

\vskip 3mm

\begin{quote}
III. {\it Глобальная жесткость}:  радиолокационное расстояние между любыми двумя точками системы отсчёта,  не меняется со временем.
\end{quote}

\vskip 3mm

Далее мы продемонстрируем, что эти три определения не эквивалентны друг другу.
Особенно неожиданным это может показаться по отношению к последним двум определениям.
В основе процедуры, дающей $\gamma_{ij}$ лежит радиолокационное измерение
расстояния между двумя бесконечно близкими точками. Однако, оказывается, что из его постоянства, вообще говоря,
не следует глобальной жесткости системы отсчёта.  То есть, бесконечно малые радиолокационные расстояния
могут быть постоянными, и при этом расстояние между удалёнными точками системы отсчёта может изменяться со временем.

\newpage

$\bullet$ Исторически первым понятие жёсткости в теории относительности ввёл в 1909г. Макс Борн  \cite{Born1909}.
Он рассматривал некоторое тело, каждая точка которого однозначно характеризуется (нумеруется)
тремя координатами $x^i=\{x,y,z\}$  и в лабораторной системе отсчёта
движется по траектории $X^\alpha=X^\alpha(\tau,x^i)$, где $\tau$ -- собственное время часов,
связанных с точкой. В классической механике тело считается жестким,  если расстояние между двумя его точками, измеренное
в данный момент времени, в дальнейшем не меняется (ниже первый рисунок).
Такое определение не является релятивистски инвариантным и в любой другой системе отсчёта будет
нарушено. Поэтому Борн потребовал для жесткого тела неизменности бесконечно малого расстояния в 4-пространстве
в гиперплоскости, ортогональной траекториям двух соседних точек (ниже второй рисунок):

\vskip 3mm
\begin{center}
\includegraphics{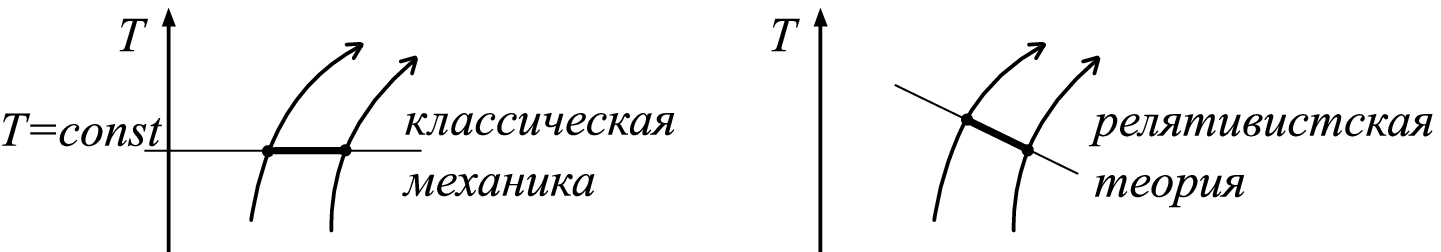}
\end{center}
\vskip 1mm

Борн использовал достаточно громоздкие нековариантные обозначения,  поэтому в приложении
мы повторим его рассуждения в существенно более компактном виде.
Мы покажем,  что критерий жёсткости Борна совпадает со вторым типом жесткости
(локальная жёсткость: $\gamma_{ij}=const$) для частного случая преобразований (\ref{transf_gen}),
в которых координатное время $t$ является собственным временем $\tau$ точки с координатами $\{x,y,z\}$.
В этой же работе Борн впервые фактически записал преобразования,
определяющие жесткую равноускоренную систему отсчёта (раздел \ref{sec_rigid_acsel}),
которую в дальнейшем активно использовал Мёллер \cite{Myelller1987}.

Отметим,  что обсуждая жесткость,  мы,  следуя Борну,  на самом деле имеем ввиду {\it кинематическую жесткость}
системы отсчёта и связанных с ней тел.
Это означает,  что эффекты деформации и соответствующих сил, действующих внутри ``твёрдого тела'',
выходят за рамки нашего рассмотрения. Жёсткая система отсчёта,
представляется как совокупность точек, расстояние между которыми при движении системы в том или ином
смысле остаётся неизменным.
С каждой точкой связан ``наблюдатель'',  имеющий часы и линейку.
Такие же эталоны времени и длины находятся у сопутствующего к нему наблюдателя в инерциальной системе отсчёта.

\newpage

\section{Нежесткая равноускоренная система отсчёта}

Пусть до момента времени $T=0$ частица покоилась в лабораторной системе отсчёта, имея координату $X=x$.
Если при $T\geqslant 0$ на неё начинает действовать постоянная сила, то координата частицы будет изменяться со временем
следующим образом \cite{Landau2_2003}:
\begin{equation}\label{nongostk_XTx}
X  = x+\frac{1}{a}\, \left[\sqrt{1+(aT)^2}-1\right],
\end{equation}
где $a$ -- константа, имеющая смысл собственного ускорения частицы (ускорение в сопутствующей к ней  инерциальной системе).
В лабораторной системе скорость частицы $U=dX/dT$ растёт, стремясь к скорости света,
а ускорение $W=dU/dT$ уменьшается:
\begin{equation}\label{x_t_UW_raket0}
U(T) = \frac{aT}{\sqrt{1+(aT)^2}},~~~~~~~~~~W(T)=\frac{a}{(1+(aT)^2)^{3/2}}.
\end{equation}
Собственное время часов, движущихся вместе с частицей со скоростью $U(T)$  равно ($\ash$ -- гиперболический арксинус):
\begin{equation}\label{nongostk_Tx}
\tau = \int\limits^T_0 \sqrt{1-U^2(T)}\,dT = \frac{1}{a}\,\ash(aT).
\end{equation}

Рассмотрим неинерциальную систему отсчёта точки которой движутся равноускоренно согласно (\ref{nongostk_XTx}).
В качестве координатного времени выберем собственное время  (\ref{nongostk_Tx}) часов в данной точке $t=\tau$.
В результате, связь координат и времён события наблюдаемого из лабораторной и неинерциальной систем отсчёта
имеет вид:
\begin{equation}\label{nonint_dsNonGostk_TX}
    T=\frac{1}{a}\sh(at),~~~~~X=x+\frac{1}{a}\left[\ch(at)-1\right],~~~~~Y=y.
\end{equation}
(мы опускаем координату $z$, так как она входит в выражения аналогично координате $y$).
Подставляя (\ref{nonint_dsNonGostk_TX}) в  (\ref{ds_mikowskij}),  получаем \cite{Logunov1987}:
\begin{equation}\label{nonint_dsNonGostk}
        ds^2 =dt^2 - 2\sh(at)\, dtdx - dx^2 - dy^2.
\end{equation}
Выделяя в $ds^2$ полный квадрат по $dt$ или используя формулу (\ref{nonint_physL}),
имеем следующее выражение для элемента физической длины:
\begin{equation}\label{nongostk_dl}
        \delta l^2 =\ch^2(at)\,dx^2+dy^2.
\end{equation}
Так как $\delta l^2$ зависит от времени, критерий {\it локальной жесткости} не выполняется.

\newpage

$\bullet$ Очевидно, что такая система не будет жесткой и в {\it глобальном} смысле.
Продемонстрируем это прямым расчётом. Пусть световой сигнал движется параллельно оси $x$.
Положив в (\ref{nonint_dsNonGostk})
$ds^2=0$ и,  выделяя полный квадрат,  получаем:
\begin{equation}\label{nonint_xt_light}
   \frac{dx}{dt} =  \pm e^{\mp at}~~~~~~~=>~~~~~~~~x(t) = const - \frac{e^{\mp at}}{a},
\end{equation}
где $const$ -- константа интегрирования, и знак минус соответствует увеличению координаты $x$, а
плюс -- уменьшению.
Пусть наблюдатель, находящийся в начале координат $x=0$,  измеряет радиолокационное расстояние до точки с координатой $x>0$.
В момент времени $t_1$ он отправляет световой сигнал, который достигает в момент времени $t$ точку  $x$,
где отражается и возвращается обратно в момент времени $t_2$.
Для определения константы в (\ref{nonint_xt_light}) при движении сигнала от наблюдателя мы выберем начальное
условие $x(t_1)=0$, и конечное условие $x(t_2)=0$ при движении в обратную сторону:
\begin{center}
\parbox{8cm}{
\includegraphics{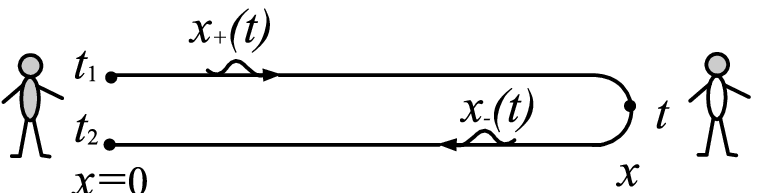}
}
\parbox{6cm}{
$$
\begin{array}{l}
\displaystyle x_+(t) = \frac{1}{a}\left(e^{-at_1}-e^{-at}\right),\\[3mm]
\displaystyle x_-(t) = \frac{1}{a}\left(\,e^{at_2}\,-\,e^{at}\,\right).
\end{array}
$$
}
\end{center}
В точке отражения $x_+(t)=x_-(t)=x$, поэтому:
$e^{-at_1}-e^{-at}=a x$ и
$e^{at_2} - e^{at} = a x.$
Исключая $t$,  находим радиолокационное расстояние:
\begin{equation}\label{nonin_t12_intx}
l=\frac{t_2-t_1}{2}  = \frac{1}{2a}\,
\ln\left[1+a x\,\frac{2\ch(at_1)-a x}{1-ax\,e^{at_1}}\right] \approx x\,\ch(at_1),
\end{equation}
где учтено, что $g_{00}=1$, поэтому собственное время равно $t$.
Приближенное равенство записано для малых $a x\ll 1$ (при фиксированном $t_1$) и соответствует физической длине (\ref{nongostk_dl}).
Из (\ref{nonin_t12_intx}) следует, что расстояние к фиксированной
точке такой равноускоренной системы меняется со временем (зависит от  $t_1$).

\vskip 2mm

$\bullet$ Относительность одновременности также приводит к нарушению {\it сопутствующей жесткости}.
Действительно, все точки неинерциальной системы  движутся синхронно относительно
лабораторной системы отсчёта $S_0:\,(T,\,X)$,
имея в данный момент времени одинаковые скорости. Однако относительно инерциальной системы $S'_0:\,(T',\,X')$,
в которой одна из точек в момент времени $T'$ неподвижна, в силу относительности одновременности,
остальные будут иметь отличные от нуля скорости.

\newpage

\section{Жесткая равноускоренная система отсчёта} \label{sec_rigid_acsel}

Рассмотрим теперь неинерциальную систему,  точки которой движутся относительно лабораторной
системы с различными собственными ускорениями и потребуем выполнения критерия {\it сопутствующей жесткости}.
Пусть произвольная точка $x$ движется по траектории:
\begin{equation}\label{gostk_XTx}
X = x + \frac{1}{a_x}\, \left[\sqrt{1+(a_xT)^2}-1\right],
\end{equation}
где $a_x$ -- константы,  различные для разных $x$ и по-прежнему $X(0)=x$.
Избавимся в (\ref{gostk_XTx}) от корня, переписав в следующем виде:
\begin{equation}\label{gostk_XTx2}
(1-a_x x)^2 + 2a_x\,(1-a_x x)\,X = 1+a^2_x\,(T^2-X^2).
\end{equation}
Подставим в это уравнение преобразования Лоренца между лабораторной системой $S_0$
и инерциальной системой $S'_0:\,\{T',\,X'\}$, движущейся относительно $S_0$ с постоянной скоростью $U_0$:
\begin{equation}\label{lorenz_transf}
T=\gamma_0\,(T'+U_0X'),~~~~~~~X=\gamma_0\,(X'+U_0T'),
\end{equation}
где $\gamma_0=1/\sqrt{1-U^2_0}$. В правой части (\ref{gostk_XTx2}) стоит инвариант, поэтому:
\begin{equation}\label{gostk_XTx3}
(1-a_x x)^2 + 2a_x\,(1-a_x x)\,\gamma_0\,(X'+U_0 T') = 1+a^2_x\,(T'^2-X'^2).
\end{equation}
Для данной точки ($x=const$) возьмём производную левой и правой части по $T'$, положив $U'=dX'/dT'=0$.
В результате,  получаем момент времени $T'$ при котором скорость точек неинерциальной системы $S$ в
инерциальной системе  $S'_0$ оказывается равной нулю:
\begin{equation}\label{gostk_XTx4}
T' = \frac{1-a_x x}{a_x}\,\gamma_0 U_0.
\end{equation}
Такое  время должно быть одинаковым для любых координат $x$ (все точки $S$ в $S'_0$ неподвижны). Это возможно,
если в (\ref{gostk_XTx4}) множитель при $\gamma_0 U_0$ не зависит от $x$:
\begin{equation}\label{sobstv_acsel}
a_x = \frac{a}{1+ax},
\end{equation}
где $a=const$ -- собственное ускорение точки $x=0$.
Подставляя $a_x$ в (\ref{gostk_XTx}),
имеем следующие траектории точек неинерциальной системы в лабораторной системе отсчёта:
\begin{equation}\label{X_x_T_gestk}
X =  \frac{1}{a}\, \left[\sqrt{(1+ax)^2+(aT)^2}-1\right].
\end{equation}
Такая система точек обладает сопутствующей жесткостью и образует жесткую
равноускоренную систему Борна-Мёллера.

\newpage

$\bullet$ Запишем преобразования между лабораторной системой $S_0:\,\{T,\,X,\,Y\}$ и жесткой равноускоренной
системой $S:\,\{t,\,x,\,y\}$:
\begin{equation}\label{nonin_TXtx_gost}
aT=(1+ax)\,\sh(at),~~~~~~~aX = (1+ax)\,\ch(at) - 1,~~~~~Y=y,
\end{equation}
где первое преобразование (определяющее координатное время $t$) выбрано таким образом, чтобы получилось
простое выражение для преобразования $X=X(t,\,x)$, после подстановки $aT$ в (\ref{X_x_T_gestk}).
Подставляя (\ref{nonin_TXtx_gost}) в интервал между событиями (\ref{ds_mikowskij}),
получаем \cite{Myelller1987}:
\begin{equation}\label{nonin_ds_gost}
ds^2 = (1+ax)^2 \,dt^2 - dx^2 - dy^2.
\end{equation}
Физическое расстояние имеет евклидовый вид $dl^2=dx^2+dy^2$, поэтому  очевидно, что критерий
{\it локальной жесткости} выполняется.

\vskip 2mm

$\bullet$ Проверим выполнимость глобальной жесткости.
Рассмотрим измерение радиолокационного расстояния, проводимое наблюдателем в точке $x=0$ до точки $x>0$ вдоль оси $x$.
Равенство нулю интервала (\ref{nonin_ds_gost}) даёт следующую траекторию светового сигнала:
\begin{equation}
\frac{dx}{dt} = \pm(1+ax)~~~~~=>~~~~~~t-t_0 = \pm \frac{1}{a}\, \ln(1+ax),
\end{equation}
где $t_0$ -- константа интегрирования и плюс соответствует увеличению координаты,  а минус -- уменьшению.
Выбирая начальное и конечное условия аналогично предыдущему разделу, имеем:
\begin{equation}
t-t_1=\frac{1}{a}\,\ln(1+ax),~~~~~~~~~~~~~t_2-t=\frac{1}{a}\,\ln(1+ax).
\end{equation}
Для наблюдателя в начале отсчёта ($x=0$) координатное и собственное времена совпадают,
поэтому радиолокационное расстояние до точки $x$ равно:
\begin{equation}\label{xpp_non_inert_xp}
l=\frac{t_2-t_1}{2}=\frac{1}{a}\, \ln\left(1+ax\right)\approx x.
\end{equation}
Заметим, что длина равна $x$ только в первом приближении по $x$.
Это на первый взгляд выглядит странным, т.к. из  выражения для бесконечно
малой физической длины $\delta l^2=dx^2+dy^2$  можно было бы ожидать $l=x$ при любых значениях $x$.
Мы обсудим причины такого несоответствия в разделе \ref{sec_geo_time_rigid}.

В любом случае, радиолокационное расстояние  $l$ постоянно,  поэтому критерий глобальной жесткости выполняется.
Следовательно, для жесткой равноускоренной системы отсчёта выполняются все три критерия жесткости.
В этом отношении она выделяется из всего многообразия неинерциальных систем отсчёта.

\newpage

\section{Вращающаяся система отсчёта}\label{sec_rot_sys}

Рассмотрим вращающийся вокруг оси $Z$ диск, лежащий в плоскости $Z=0$.
Запишем  интервал лабораторной системы (\ref{ds_mikowskij})
в цилиндрических координатах $X=R\cos\Phi$, $Y=R\sin\Phi$:
\begin{equation}\label{nonin_rot_ds_cilindr}
        ds^2 = dT^2 - dR^2 - R^2 \, d\Phi^2
\end{equation}
и выполним преобразования:
\begin{equation}\label{nonin_rot_transf_Born}
       T=t,~~~~~~R=r,~~~~\Phi = \phi + \omega t,
\end{equation}
где координаты $(r,\,\phi)$ нумеруют точки вращающейся системы, связанной с диском.
Так как $T=t$, имеем $\Phi = \phi + \omega T$, что является  траекторией некоторой точки,  вращающейся
по окружности с постоянной угловой скоростью $\omega$.
Подставляя (\ref{nonin_rot_transf_Born}) в (\ref{nonin_rot_ds_cilindr}),  имеем \cite{Landau2_2003}:
\begin{equation}\label{nonin_rot_ds_Born}
ds^2 = (1-\omega^2 r^2)\,dt^2 - 2\omega\,  r^2 \,  dt\,d\phi - dr^2 - r^2\, d\phi^2.
\end{equation}
Предполагается,  что $\omega r<1$, т.е. размеры диска ограничены и все его точки движутся
медленнее скорости света.
Элемент физической длины, соответствующий метрике (\ref{nonin_rot_ds_Born}):
\begin{equation}\label{nonin_geum_rot_dl}
\delta l^2 = dr^2 + \frac{r^2\,d\phi^2}{1-\omega^2 r^2}
\end{equation}
не зависит от времени, а,  следовательно,  равномерно вращающаяся система отсчёта
является {\it локальной жесткой}.

$\bullet$ Однако в этом случае {\it сопутствующая жесткость} во вращающейся
системе не выполняется. Действительно, пусть сопутствующая система отсчёта
движется со скоростью $\omega r_1$ и её начало в данный момент находится внизу окружности, как это изображено на первом рисунке (ось $Z$ направлена к читателю):
\begin{center}
\includegraphics{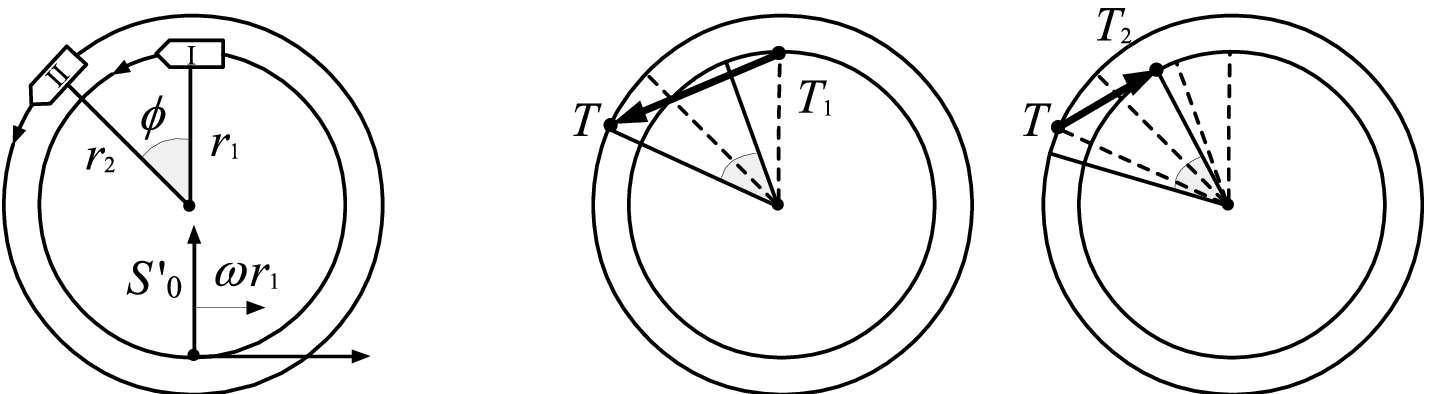}
\end{center}
Соответствующая точка вращающейся системы, совпадающая с началом сопутствующей,  имеет в последней нулевую скорость.
Однако другие точки (например, изображенные в виде космических кораблей)
имеют в сопутствующей системе отличные от нуля скорости.

\vskip 2mm

$\bullet$ Проверим теперь критерий {\it глобальной жесткости}.
Пусть  два космических корабля в лабораторной системе вращаются с одинаковой угловой скоростью
$\omega$ и находятся на различных расстояниях $r_1$ и $r_2$ от центра. Угол между прямыми, проведенными к кораблям
из центра вращения равен $\phi$.
Скорость первого корабля $\omega r_1$ определяет связь его корабельных часов с часами лабораторной системы:
\begin{equation} \label{nonint_rot_Tr1r2}
         \tau =T\sqrt{1-(\omega r_1)^2}.
\end{equation}
Это соотношение можно получить как из общего выражения для собственного времени $\delta\tau = \sqrt{g_{00}}\,dt$ и $t=T$,  так и при помощи стандартной релятивистской
формулы замедления времени.

Пусть наблюдатель на первом корабле с радиусом орбиты $r_1$
проводит измерение радиолокационного расстояния ко второму кораблю.
Вычисления проще провести в лабораторной системе отсчета.
В этой системе за время $T-T_1$ распространения сигнала в одну сторону (выше средний рисунок),  второй корабль смещается на угловое расстояние
$\omega (T-T_1)$. Поэтому длина пути сигнала находится по теореме косинусов с углом $\phi+\omega\,(T-T_1)$.
После  отражения сигнала, он движется навстречу первому кораблю и для вычисления длины траектории возвращения
сигнала за время $T_2-T$ в теореме косинусов необходимо взять угол $\phi-\omega\,(T_2-T)$.
В результате:
\begin{equation}\label{nonin_rot_T1T2_cos}
\left\{
\begin{array}{l}
r^2_1+r^2_2-2r_1r_2\cos \left[\phi+\omega\,(T-T_1)\right] = (T-T_1)^2, \\ [4mm]
r^2_1+r^2_2-2r_1r_2\cos \left[\phi-\omega\,(T_2-T)\right] = (T_2-T)^2.
\end{array}
\right.
\end{equation}
В левой части уравнений находится квадрат длины пути светового сигнала в лабораторной системе,
а в правой -- квадрат времени его движения (скорость света равна
единице).
Запишем решения этих трансцендентных уравнений относительно времён:
\begin{equation}
T-T_1 = f(\omega),~~~~~~T_2-T=f(-\omega),
\end{equation}
где $f(\omega)$ -- функция угловой скорости $\omega$, а также радиусов $r_1$, $r_2$ и угла $\phi$,
зависимость от которых опущена.
Исключая $T$ и переходя к собственному времени первого корабля (\ref{nonint_rot_Tr1r2}), имеем:
\begin{equation}\label{nonin_rot_l}
       l= \frac{\tau_2-\tau_1}{2} = \frac{1}{2}\,\sqrt{1-(\omega r_1)^2}~ [ f(\omega)+f(-\omega)].
\end{equation}
Правая часть этого соотношения не зависит от времени. Поэтому  радиолокационный эксперимент
приведет наблюдателя к выводу, что расстояние между кораблями неизменно.


\newpage

\section{Поступательно ускоренные системы Мёллера}\label{sec_Meller_vt}

До сих пор критерии локальной и глобальной жесткости совпадали.
Рассмотрим  класс неинерциальных систем,  в которых это не так.
Для этого запишем  следующие преобразования от неинерциальной системы $S:\,(t,x)$ к лабораторной системе $S_0:\,(T,X)$ \cite{Myelller1987}:
\begin{equation}\label{nonin_gesk1}
T=\gamma\, v\,x + \int\limits^t_0 \gamma\, dt,~~~~~~X=\gamma\, x+\int\limits^t_0 \gamma\, v\,dt,~~~~~~~Y=y,
\end{equation}
где $v=v(t)$ произвольная функция времени и $\gamma=\gamma(t)=1/\sqrt{1-v^2}$.
Если скорость $v$ постоянна,
то (\ref{nonin_gesk1}) приводят к преобразованиям Лоренца.
В общем же случае имеем:
\begin{equation}\label{meller_dT_dX}
dT = \gamma\, (dt+v\,dx) + \gamma^3\,\dot{v}\,x\,  dt,~~~~~dX = \gamma\, (dx+v\,dt)+\gamma^3\,\,v\dot{v}\, x\,dt,
\end{equation}
где учтено, что $d\gamma/dt = \gamma^3 v\dot{v}$,
$d(\gamma v)/dt=\gamma^3\,\dot{v}$ и точка -- производная по времени $t$.
Из (\ref{meller_dT_dX}) следует, что фиксированная точка $x=const$  ($dx=0$)
движется относительно лабораторной системы отсчета со скоростью $U(T) =  dX/dT  = v(t).$
Хотя эта скорость не зависит от координаты $x$, это не означает, что в лабораторной системе
скорость всех точек неинерциальной системы одинакова.
Дело в том, что $v(t)$ -- это скорость, вычисляемая в момент времени $t$ по часам неинерциальной системы отсчёта.
Чтобы получить скорость $U(T)$ данной точки $x$  относительно лабораторной системы,
в функции $v(t)$ необходимо перейти от координатного времени $t$ ко времени $T$,
которое находится из первого преобразования (\ref{nonin_gesk1}).
При этом получится некоторая функция $t=t(T,\,x)$ и  в данный момент времени $T$
по лабораторным часам скорости точек с различными координатами $x$ будут различны.

Подставляя  дифференциалы в интервал (\ref{ds_mikowskij}), имеем:
\begin{equation}\label{nonin_gesk2}
ds^2=\left[1+w(t)\,  x\right]^2\, dt^2-dx^2-dy^2,
\end{equation}
где $w(t)=\gamma^2\dot{v}$.
Если $w(t)=a=const$,  мы возвращаемся к жесткой равноускоренной системе отсчета,
для которой $v(t) = \th(at)$.
В этом случае произвольная точка системы движется релятивистски равноускоренно
с собственным ускорением (\ref{sobstv_acsel}).

Очевидно, что для метрики (\ref{nonin_gesk2}) физическая длина является евклидовой $\delta l^2=dx^2+dy^2$.
На этом основании Мёллер назвал класс неинерциальных систем отсчёта (\ref{nonin_gesk1}) жесткими.
Разберёмся, однако, выполняется ли для них критерий глобальной жесткости.

\newpage

$\bullet$ Равенство нулю интервала (\ref{nonin_gesk2}) приводит к следующему дифференциальному уравнению:
\begin{equation}\label{moller_light}
 \frac{dx}{dt} = \pm(1+w\,x).
\end{equation}
Пусть световой сигнал отправляется в момент времени $t_1$ из начала системы отсчёта $x=0$.
В момент времени $t_2$ он туда
возвращается,  отразившись от точки с координатой $x>0$.
Рассмотрим движение в сторону возрастания координаты (знак плюс).
Так как $x(t_1)=0$, из (\ref{moller_light}) следует, что в момент времени $t=t_1$:
\begin{equation}\label{dxdt_meller}
 \frac{dx}{dt}\Bigr|_{t=t_1} = 1,~~~~~~~\frac{d^2x}{dt^2}\Bigr|_{t=t_1} = (\dot{w} x + \frac{dx}{dt}\,w)_{t=t_1}=w(t_1),
\end{equation}
где вторая производная получена дифференцированием (\ref{moller_light}). Поэтому
траектория удаляющегося сигнала имеет вид:
\begin{equation}
x_+(t)\approx (t-t_1) + w(t_1)\,\frac{(t-t_1)^2}{2}+...
\end{equation}
Аналогично находится траектория приближающегося к началу отсчёта сигнала $x(t_2)=0$, соответствующая  в (\ref{moller_light}) знаку минус:
\begin{equation}
x_-(t)\approx (t_2-t) + w(t_2)\,\frac{(t_2-t)^2}{2}+...
\end{equation}
При отражении эти две траектории совпадают: $x_+(t)=x_{-}(t)=x$.
Решая квадратные уравнения
относительно $t-t_1>0$ и $t_2-t>0$ и складывая решения, получаем:
\begin{equation}
t_2-t_1 \approx \frac{\sqrt{1+2w_1\,  x}-1}{w_1}+\frac{\sqrt{1+2w_2 \,x}-1}{w_2},
\end{equation}
где $w_1=w(t_1)$ и $w_2=w(t_2)$. При малом интервале времени $t_2-t_1$ координата точки отражения является величиной того же порядка
малости. Поэтому  разложим  корень до малых $x^2$ включительно:
\begin{equation}
l=\frac{t_2-t_1}{2} \approx x - \frac{w_1}{2}\,x^2,
\end{equation}
где член $(w_1+w_2)x^2$ с точностью до второго порядка малости заменен на $2 w_1 x^2$.
Так как для наблюдателя в начале системы отсчёта $x=0$,  собственное время совпадает с координатным и
полученное выражение является радиолокационным расстоянием к точке с координатой $x$.
В первом порядке малости это расстояние постоянно, что отражено в постоянстве
физической длины $\delta l^2=dx^2+dy^2$. Однако уже следующее приближение по $x$ оказывается
зависящим от времени посылки сигнала, если только величина  $w(t)$ не является константой.

\newpage

\section{Жесткость, время и геометрия}\label{sec_geo_time_rigid}

Как мы видели,  постоянство бесконечно малого радиолокационного расстояния $\delta l^2 = \gamma_{ij}\,dx^idx^j$,
вообще говоря,  не гарантирует,  что конечное радиолокационное расстояние между двумя точками будет постоянным
(локальная жёсткость не влечёт за собой глобальной жёсткости).
Это свойство неинерциальных систем отсчёта тесно связано с другой, отмеченной выше,  особенностью.
В жесткой равноускоренной системе радиолокационное расстояние равно $l=\ln(1+ax)/a$.
В то же время $\delta l^2=dx^2+dy^2$ и при движении вдоль оси $x$ мы,  на первый взгляд,  должны были бы иметь $l=x$.

Причина этих расхождений кроется в измерительном смысле физической длины $\delta l^2 = \gamma_{ij}\,dx^idx^j$.
Её получает наблюдатель, измеряя {\it время} распространения светового сигнала в обе стороны к бесконечно близкой к нему точке.
Суммирование малых элементов $\delta l$ вдоль некоторой кривой
подразумевает, что вдоль этой кривой расположено множество таких наблюдателей, каждый из которых получает своё значение $\delta l$.
Однако  время для разных наблюдателей в неинерциальной системе отсчёта, в общем случае,
течёт различным образом. Поэтому  сумма измерений радиолокационных расстояний в которых используются часы,
расположенные в различных точках, отличается от единственного измерения такого же расстояния,
проведенного одним наблюдателем по одним часам.

Хорошей иллюстрацией этого утверждения является вращающаяся система отсчёта. Для наблюдателей,
находящихся на одинаковом расстоянии от центра,  темп хода часов одинаков.
В разделе \ref{sec_rot_sys} мы рассматривали движение светового сигнала по геодезической.
В принципе,  можно измерять длину любой линии,  вдоль которой распространяется свет.
Экспериментально такая линия может быть организована при помощи световода или системы зеркал.
Пусть сигнал движется по окружности ($r=const$) от точки $\phi=0$,  до точки $\phi>0$
и обратно. Равенство нулю интервала (\ref{nonin_rot_ds_Born}) приводит к уравнению:
\begin{equation}
\frac{d\phi}{dt}  = \pm \frac{1}{r} - \omega.
\end{equation}
Повторяя рассуждения предыдущих разделов, имеем:
\begin{equation}
l =  \sqrt{g_{00}}~\frac{t_2-t_1}{2} = \frac{r\phi}{\sqrt{1-(\omega r)^2}}.
\end{equation}
Такое же расстояние мы получим,  интегрируя выражение для физической длины (\ref{nonin_geum_rot_dl})
во вращающейся системе при $r=const$.

\newpage

Иная ситуация будет при движении светового сигнала вдоль радиуса ($\phi=const$).
В этом случае уравнение его движения
\begin{equation}
\frac{dr}{dt}  = \pm \sqrt{1-(\omega r)^2}
\end{equation}
приводит к следующему радиолокационному расстоянию для наблюдателя,  расположенного в центре вращения:
\begin{equation}
 l = \frac{1}{\omega}\, \arcsin(\omega r).
\end{equation}
Это выражение уже отличается от $l=r$, которое следует при интегрировании (\ref{nonin_geum_rot_dl})
вдоль линии $\phi=const$.

Таким образом, одинаковый темп течения времени вдоль траектории светового сигнала приводит к совпадению
результата единичного радиолокационного измерения расстояния и суммы измерений бесконечно малых расстояний.
Если же темп течения времени вдоль траектории различен, то результаты измерений будут отличаться.

В связи с этим отметим ещё один момент.
Отклонение физической длины $\delta l^2 = \gamma_{ij}\,dx^idx^j$ от евклидового выражения,
обычно интерпретируется как неевклидовость 3-прос\-тран\-ства в неинерциальной системе отсчёта.
Например, 3-прос\-тран\-ство  вращающейся системы отсчёта (\ref{nonin_geum_rot_dl}) с метрикой $\gamma_{ij}$ имеет отрицательную кривизну.
Стоит, однако иметь ввиду, что подобная неевклидовость существенно отличается от неевклидовости обычных
искривлённых пространств.
В геометрии нет времени.  Длина линии должна равняться сумме длин её бесконечно малых элементов.
Однако оба эти утверждения не выполняются в неинерциальных системах отсчёта.
Поэтому,  рассмотрение геометрических свойств пространства, например,  с метрикой  (\ref{nonin_geum_rot_dl})
несколько формально. К примеру, физическая длина в жесткой равноускоренной системе
евклидова: $\delta l^2=dx^2+dy^2$. Эта длина получена в результате анализа распространения света на бесконечно малое расстояние.
Однако,  в таком евклидовом пространстве тот же свет, распространясь на конечные расстояния,
движется не по прямым, а по искривлённым линиям.

За геометрическими свойствами метрики $\gamma_{ij}$ необходимо видеть множество наблюдателей,
использующих различные часы для измерения радиолокационных расстояний в своих непосредственных окрестностях.
Геометрия 3-пространства неинерциальной системы, основанная на $\gamma_{ij}$,
является геометрией, объединяющей такие бесконечно малые локальные измерения.

\newpage

\section{Глобально жёсткие поступательно движущиеся системы}

Из рассмотренных выше неинерциальных систем отсчёта только жесткая равноускоренная система
обладала жёсткостью с точки зрения всех критериев жёсткости.
Выясним,  возможно ли определить отличную от неё неинерциальную систему,
обладающую аналогичным свойством в классе поступательно движущихся систем отсчёта.
В общем случае метрика системы, движущейся относительно лабораторной системы вдоль оси $X$,  имеет вид:
\begin{equation}
ds^2 = g_{00}\, dt^2 + 2 g_{01}\, dtdx + g_{11}\, dx^2.
\end{equation}
При помощи переопределения координатного времени $t=t(t',\,x')$ этот интервал всегда можно диагонализовать,
обнулив метрический коэффициент $g_{01}$.
Если после этого  $g_{11}$ зависит от времени, то такая система не обладает
локальной жесткостью,  а,  следовательно, не является жесткой и в глобальном смысле.
Поэтому  жёсткая система отсчёта должна иметь $g_{11}$,  не зависящий от времени.
При помощи изменения способа нумерации точек системы $x=x(x')$ этот коэффициент можно сделать
единичным.
Следовательно,  без потери общности метрика локально жёсткой системы отсчёта записывается в следующем виде:
\begin{equation}\label{ds_f_1}
ds^2 = f^2(t,x)\, dt^2 - dx^2.
\end{equation}

Мы ограничиваемся анализом неинерциальных систем в рамках специальной теории относительности,
в которой пространство является псевдоевклидовым и имеет нулевой тензор кривизны:
\begin{equation}
R^{\alpha}_{~\beta,\,\mu\nu} =
 \partial_\mu \Gamma^\alpha_{\beta \nu}-\partial_\nu \Gamma^\alpha_{\beta \mu}
+\Gamma^\alpha_{\sigma \mu}\Gamma^\sigma_{\beta  \nu}-\Gamma^\alpha_{\sigma \nu}\Gamma^\sigma_{\beta \mu} = 0.
\end{equation}
Выясним при каких функциях $f=f(t,x)$ это происходит.
Ненулевые символы Кристоффеля, соответствующие метрике (\ref{ds_f_1}) равны:
\begin{equation}\label{ds_f_G}
\Gamma^t_{tx}=\Gamma^t_{tx} = \frac{\partial_x f}{f},~~~~~~\Gamma^t_{tt} = \frac{\partial_t f}{f},~~~~~~\Gamma^x_{tt}=f\, \partial_x f,
\end{equation}
где $\partial_x f$ -- частная производная функции $f=f(t,x)$ по $x$,  а $\partial_t f$ -- по $t$
и $\Gamma^t_{tx}=\Gamma^0_{01}$,  и т.д.
В пространстве размерности $1+1$, с учётом свойств симметрии, единственной нетривиальной компонентой тензора кривизны
будет компонента:
\begin{equation}
R^{t}_{~x,\,tx} =
 -\partial_x \Gamma^t_{x t}-\Gamma^t_{t x}\Gamma^t_{x t},
\end{equation}
где сразу опущены нулевые символы Кристоффеля.

\newpage

Это выражение равно нулю, если выполняется
уравнение:
\begin{equation}
 \partial_x \Gamma = -\Gamma^2~~~~~~~=>~~~~~~~~~\Gamma=\frac{1}{x+\alpha(t)},
\end{equation}
где $\Gamma=\Gamma^t_{tx}$ и $\alpha(t)$ -- произвольная функция времени.
Интегрируя ещё раз $\Gamma=\partial_xf/f$,  получаем:
\begin{equation}
 f(t,x) = \beta(t)\,(\,x+\alpha(t)\,).
\end{equation}
С учётом оставшегося произвола в выборе преобразования координатного времени $t=t(t')$,
эта функция совпадает с метрическим коэффициентом $g_{00}$
поступательно движущихся неинерциальных систем Мёллера из раздела \ref{sec_Meller_vt}.  Они являются жесткими
в глобальном смысле только,  если $g_{00}=f^2$  не зависит от времени.

Таким образом, единственной поступательно движущейся неинерциальной системой отсчёта,
в которой одновременно выполняется критерии локальной и глобальной жесткости,  является
жесткая равноускоренная система отсчёта Борна-Мёллера,  рассмотренная в разделе \ref{sec_rigid_acsel}.


\section{Заключение}

Проведенный анализ показывает, что локальная жесткость системы отсчёта, вообще говоря,
не влечет за собой жесткости глобальной. Не смотря на то, что в основе каждого критерия
лежит радиолокационный эксперимент, смысл его организации отличается.
При глобальном измерении расстояния наблюдатель отправляет световой сигнал на конечное
расстояние. При локальном измерении это делается для бесконечно близкой к наблюдателю точке.
При помощи последовательности измерений бесконечно малых расстояний можно измерить и конечное расстояние.
Однако такой  эксперимент будет проводить не один наблюдатель,  а множество наблюдателей,
расположенных между двумя удалёнными точками. Эти наблюдатели имеют различный темп течения собственного
времени. В результате,  сумма их последовательных измерений, в общем случае,
будет отличаться от единственного измерения радиолокационного расстояния, проведенного одним наблюдателем.

Автор благодарит Орлянского О.Ю. за полезные обсуждения вопросов, затронутых в статье.

\newpage
\section{Приложение: жёсткость по Борну}

Приведём в ковариантных обозначениях вывод условия жёсткости Борна \cite{Born1909}.
Вместо жёсткой неинерциальной системы,  следуя Борну, будем  говорить о жёстком теле.
Запишем траекторию произвольной точки такого тела относительно лабораторной системы $S_0:\,\{T,\,X,\,Y,\,Z\}$:
\begin{equation}\label{borm_traject}
X^\alpha=X^\alpha(\tau,\,  x^1,x^2,\,x^3).
\end{equation}
При этом,  $x^i$ -- это координаты, однозначно определяющие фиксированную точку тела,  а $\tau$ -- её собственное время.
Ниже мы используем безындексную запись в которой прямым шрифтом будут обозначаться 4-векторы
(скалярное произведение $A^\alpha B_\alpha$ имеет вид $\mathrm{A}\cdot\mathrm{B}$ и т.д.).

Интервал вдоль траектории движения точки $x^i=const$ совпадает с изменением её собственного времени:
\begin{equation}
d\tau^2 =  dX^\alpha dX_\alpha = (\partial_0 X^\alpha)(\partial_0 X_\alpha)\,  d\tau^2\equiv (\partial_0 \mathrm{X})^2\,  d\tau^2.
\end{equation}
Поэтому:
\begin{equation}\label{born_tau_def}
 (\partial_0 \mathrm{X})^2=1,
\end{equation}
где $\partial_0=\partial/\partial\tau$ -- производная по собственному времени при постоянстве $x^i=const$.
Рассмотрим две соседние точки с координатами $x^i$ и $x^i+dx^i$.
Положение первой точки соответствует моменту собственного времени $\tau$, а второй: $\tau+d\tau$.
Расстояние между этими точками в пространстве Минковского  определяется 4-вектором:
\begin{equation}\label{born_dX}
d\mathrm{X} = \partial_0\mathrm{X}\,d\tau + \partial_i\mathrm{X}\, dx^i,
\end{equation}
где $\partial_i=\partial/\partial x^i$.
Для определения значения $d\tau$, потребуем,  чтобы этот вектор был ортогонален
к 4-вектору $\partial_0\mathrm{X}$, касательному к траектории при изменении собственного времени (первое уравнение):
\begin{center}
\parbox{8cm}{
\includegraphics{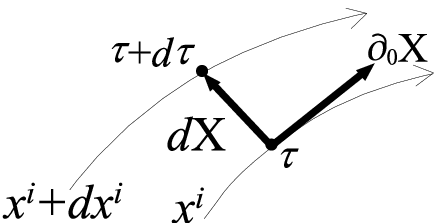}
}
\parbox{6.5cm}{
\begin{equation}
\left\{
\begin{array}{l}
d\mathrm{X}\cdot\partial_0\mathrm{X}= 0,\\[3mm]
(d \mathrm{X})^2 = const.
\end{array}
\right.
\end{equation}
}
\end{center}
\vskip 2mm
\noindent
При таком выборе тело,  по определению,  считается {\it локально жёстким},
если длина вектора $d \mathrm{X}$ не меняется со временем (второе уравнение).

\newpage

Из соотношения ортогональности и (\ref{born_dX}), (\ref{born_tau_def}) получаем
\begin{equation}
d\tau = -(\partial_0\mathrm{X}\cdot\partial_i\mathrm{X})\, dx^i.
\end{equation}
Подставляя это значение в квадрат расстояния (\ref{born_dX}) между точками
\begin{equation}
(d \mathrm{X})^2 = d\tau^2 + 2(\partial_0\mathrm{X}\cdot\partial_i\mathrm{X})\, d\tau dx^i +(\partial_i\mathrm{X}\cdot\partial_j\mathrm{X})\, dx^idx^j,
\end{equation}
имеем:
\begin{equation}
(d\mathrm{X})^2 = \bigl\{ \partial_i\mathrm{X}\cdot\partial_j\mathrm{X} -  (\partial_0\mathrm{X}\cdot\partial_i\mathrm{X})
(\partial_0\mathrm{X}\cdot\partial_j\mathrm{X})\bigr\}\,   dx^idx^j.
\end{equation}
Первое слагаемое в фигурных скобках -- это $g_{ij}$,  а второе -- произведение $g_{0i}g_{0j}$.
Действительно,  интервал равен:
$$
ds^2 = (d\mathrm{X})^2 = (\partial_\alpha\mathrm{X}\cdot \partial_\beta \mathrm{X})\, dx^\alpha dx^\beta = g_{\alpha\beta}\, dx^\alpha
dx^\beta,
$$
поэтому  метрические коэффициенты $g_{\alpha\beta}$ неинерциальной системы,  связанной с телом,  равны $\partial_\alpha\mathrm{X}\cdot \partial_\beta \mathrm{X}$.
При этом, так как в преобразованиях (\ref{borm_traject}) $\tau$ -- собственное
время, то $g_{00}=(\partial_0\mathrm{X})^2=1$.

Таким образом,  критерий жёсткости Борна эквивалентен постоянству тензора $\gamma_{ij}$,
определяющего физическую длину (\ref{nonint_physL}).

\vskip 5mm

\begin{flushright}
2013-06-20
\end{flushright}

\newpage

\begin{center}
{\bf \huge Are rigid non-inertial frames of reference really rigid?}

\vskip 10mm

{\huge S.\,S.\,Stepanov} \\
~\\

\vskip 5mm

\parbox{14cm}{
\large
In this paper the notion of the rigid frame of reference within special relativity is analysed.
Three definitions of rigidity are formulated.
By using several examples of non-inertial frames, it is shown that these definitions are not equivalent.
It is also shown that so called M\"oller rigid non-inertial frames are locally rigid, but do not exhibit global rigidity.
The physical meaning of this phenomenon is discussed, as well as its relation to the non-Euclidean nature of space
in non-inertial frames of reference.
}

\end{center}

\end{document}